# Fractional Dynamics of Natural Growth and Memory Effect in Economics

**Valentina V. Tarasova**
Higher School of Business, Lomonosov Moscow State University,
Moscow 119991, Russia; E-mail: v.v.tarasova@mail.ru

**Vasily E. Tarasov**
Skobeltsyn Institute of Nuclear Physics, Lomonosov Moscow State University,
Moscow 119991, Russia; E-mail: v.v.tarasov@bk.ru; tarasov@theory.sinp.msu.ru

**Abstract:** A generalization of the economic model of natural growth, which takes into account the power-law memory effect, is suggested. The memory effect means the dependence of the process not only on the current state of the process, but also on the history of changes of this process in the past. For the mathematical description of the economic process with power-law memory we used the theory of derivatives of non-integer order and fractional-order differential equation. We propose equations take into account the effects of memory with one-parameter power-law damping. Solutions of these fractional differential equations are suggested. We proved that the growth and downturn of output depend on the memory effects. We demonstrate that the memory effect can lead to decrease of output instead of its growth, which is described by model without memory effect. Memory effect can lead to increase of output, rather than decrease, which is described by model without memory effect.

## 1. Introduction

Natural growth models are widely used in physics, chemistry, biology and economics. The economic models of natural growth are described by equations in which the marginal output (the growth rate of output) is directly proportional to income. A more realistic model is considered to be natural growth, which marginal output depends on the profit instead of the income.

We first describe the simplest version of the economic model of natural growth that does not take into account the effects of time delay (lag) [1] and the memory effects [2, 3, 4, 5]. Let $Y(t)$ be a function that described the value of output at time t. We use the approximation of the unsaturated market, implying that all manufactured products are sold. We also assume that the volume of sales is not large, and, therefore, does not affect the price of the goods, that is, the price P>0 is assumed a constant. Let $I(t)$ be a function that describes the net investment, i.e. the investment that is used to the expansion of production. In the model of natural growth is assumed that the marginal income (d (P·Y(t))/dt) and the rate of output ((dY(t))/dt) are directly proportional to the value of the net investment, and we can use the accelerator equation

$$\frac{dY(t)}{dt} = \frac{1}{v} \cdot I(t), \qquad (1)$$

where v is a positive constant called the investment ratio and characterizes the accelerator power, 1/v - marginal productivity of capital (rate of acceleration), and dY(t)/dt is the first order derivative of the function Y(t) with respect to time. Assuming that the value of the net investment is a fixed part of the profit, which is equal to the difference between the income of P·Y(t) and the costs C(t), we have the equation

$$I(t) = m \cdot (P \cdot Y(t) - C(t)), \qquad (2)$$

where m is the rate of net investment (0<m<1), i.e. the share of profit, which is spent on the net investment. We assume that the costs C(t) are linearly dependent on the output Y(t), such that we have the equation

$$C(t) = a \cdot Y(t) + b, \qquad (3)$$

where a is the marginal costs, and b is the independent costs, i.e. the part of the cost, which does not depend on the value of output. Substituting expressions (2) and (3) into equation (1), we obtain

$$\frac{dY(t)}{dt} - \frac{m \cdot (P-a)}{v} \cdot Y(t) = -\frac{m \cdot b}{v}. \qquad (4)$$

Differential equation (4) describes the economic model of natural growth without memory and lag.

The general solution of differential equation (4) has the form

$$Y(t) = \frac{b}{(P-a)} + c \cdot \exp\left(\frac{m \cdot (P-a)}{v} \cdot t\right), \qquad (5)$$

where c is a constant. Using the initial value Y(0) of function (5) at t = 0, we obtain c=Y(0)-b/(P-a). As a result, we have the solution

$$Y(t) = \frac{b}{(P-a)}\left(1 - \exp\left(\frac{m \cdot (P-a)}{v} \cdot t\right)\right) + Y(0) \cdot \exp\left(\frac{m \cdot (P-a)}{v} \cdot t\right). \qquad (6)$$

Solution (6) of equation (4) describes the dynamics of output within the natural growth model without memory effects.

In the model of natural growth, which is described by equation (4), is supposed to perform the accelerator equation (1) that connects the net investment and the marginal value of output. Equations (1) and (4) contain only the first-order derivative with respect to time. It is known that the derivative of the first order is determined by the properties of differentiable functions of time only in infinitely small neighborhood of the time point. Because of this, the natural growth model (4) involves an instantaneous change of output speed when changing the net investment. This means that the model of natural growth (4) neglects the effects of memory and delay.

In this paper, the economic model of natural growth, which is described by equation (4), is generalized by taking into account the power-law dynamic memory. The memory is considered as a property that describes a dependence of the variables not only on the current state of the process, but also on the changes of these variables in the past. To describe such economic processes, we used the derivatives of non-integer order and fractional-order differential equation. We propose fractional differential equations of the economic natural growth with memory and corresponding solutions. We proved that the economic growth and downturn depend on the memory effects: (a) the memory effect can lead to downturn instead of growth, which is described by model without memory; (b) the memory effect can lead to growth, rather than downturn, which is described by model without memory effect.

## 2. Memory Effects by Fractional Derivatives and Integrals

We use the concept of dynamic memory to describe the economic processes, analogous to the use of this concept in physics [3, p. 394-395]. Dynamic memory is an average characteristic of process that describes the dependence of the process state at present time of the process state in the past [2, 3, 4, 5].

Let us describe the concept of memory for economics. Economic process with memory is a process, where economic parameters and factors (endogenous and exogenous variables) at a given time depend not only on their values at this time. This process also depends on the values of the endogenous and exogenous variables at previous times [6, 7, 8, 9, 10, 11, 12].

To consider memory effects in natural growth model, we assume that the value of output Y(t) at time t depends not only on the net investment I(t) at the same time point, but also depends on the changes I($\tau$) on a finite time interval [0, t]. This is due to the fact that economic agents can remember the previous changes of investments I(t) and the impact of these changes on the value of output Y(t). To describe this effect (memory effect) the dependence of output from net investment, we can use the equation

$$Y(t) = \int_0^t M(t - \tau) \cdot I(\tau) d\tau, \qquad (7)$$

where M(t) is the memory function that allows you to take into account the memory of the norms of the net investment. If the function M(t) is expressed by the Dirac delta-function (M(t) = M·$\delta$(t)), then equation (7) becomes the standard equation of multiplier Y(t)=M·I(t). If the function M(t) has the form M(t) = M·$\delta$(t-T), then equation (7) becomes the equation of multiplier with fixed-time delay Y(t)=M·I(t-T), [1, p. 25]. If the normalization condition $\int_0^t M(t - \tau) d\tau = 1$ holds for the function M(t), then equation (7) is often interpreted as the equation with the continuously distributed lag and the function is called the weighting function. In this case, also speak of the complete memory [3, p. 395], since the process passes through all states continuously without any loss. If we assume the power law fading of memory, we can use the memory function

$$M(t - \tau) = \frac{1}{\Gamma(\alpha)} \frac{M}{(t-\tau)^{1-\alpha}}, \qquad (8)$$

where $\alpha > 0$, $t > \tau$, and $\Gamma(\alpha)$ is the Gamma function. In order to have the correct dimensions of economic quantities we will use the dimensionless time variable t. In this case, the equation (7) takes the form

$$Y(t) = M \cdot (I_{0+}^\alpha I)(t), \qquad (9)$$

where $I_{0+}^\alpha$ is the left-sided Riemann-Liouville fractional integral of order $\alpha > 0$ the variable t that is defined by the equation

$$(I_{0+}^\alpha I)(t) := \frac{1}{\Gamma(\alpha)} \int_0^t \frac{I(\tau) d\tau}{(t-\tau)^{1-\alpha}}, \qquad (10)$$

where $0 < \tau < t$, and the function I($\tau$) is measurable on the interval (0,t) and satisfies $\int_0^t |I(\tau)| d\tau < \infty$. Equation (9) describes the multiplier of non-integer order [9], and M is a multiplier coefficient of this multiplier. This allows us to use the fractional calculus and fractional differential equations [13, 14, 15, 16] to describe economic processes with memory.

In order to express the function of net investment I(t) through the function Y(t), which describes the output, we act on equation (9) by the Caputo fractional derivative of order $\alpha > 0$ that is defined by the equation

$$(D_{0+}^\alpha Y)(t) := \frac{1}{\Gamma(n-\alpha)} \int_0^t \frac{Y^{(n)}(\tau) d\tau}{(t-\tau)^{\alpha-n+1}}, \qquad (11)$$

where $Y^{(n)}(\tau)$ is the derivative of integer order n: = $[\alpha]$ +1 of the function $Y(\tau)$ with respect to $\tau$ such that $0 < \tau < t$. Here function $Y(\tau)$ must have the derivatives of integer orders up to the (n-1)th order, which are absolutely continuous functions on the interval [0, t]. The economic interpretation of this fractional-order derivative has been suggested in [17].

The action of derivative (11) on equation (9) gives the expression

$$(D_{0+}^\alpha Y)(t) = M \cdot (D_{0+}^\alpha I_{0+}^\alpha I)(t). \tag{12}$$

It is known that the Caputo derivative is inverse to the Riemann-Liouville integral [15, p. 95], and for any continuous function f (t) $\in$C [0, t] the identity

$$(D_{0+}^\alpha I_{0+}^\alpha f)(t) = f(t) \tag{13}$$

holds for any $\alpha > 0$, where $I_{0+}^\alpha$ is the left-sided Riemann-Liouville fractional integral (10) and $D_{0+}^\alpha$ is the left-sided Caputo fractional derivative (11).

Using identity (13), equation (12) can be written as

$$(D_{0+}^\alpha Y)(t) = \frac{1}{v} \cdot I(t), \tag{14}$$

where v=1/M. As a result, the multiplier with memory (9) can be represented in the form of the accelerator with memory (14), where the coefficient of the accelerator is the inverse of the coefficient multiplier (9).

Accelerator equation (14) contains the standard equation of the accelerator and the multiplier, as special cases. To prove this, consider equation (14) for $\alpha = 0$ and $\alpha = 1$. Using property $(D_{0+}^1 X)(t) = X^{(1)}(t)$ of the Caputo fractional derivative [15, p. 79], equation (14) with $\alpha = 1$ gives equation (1) that describes the standard accelerator. Using the property $(D_{0+}^0 Y)(t) = Y(t)$, equation (14) with $\alpha = 0$ can be written as $I(t) = v \cdot Y(t)$, that is a standard equation of multiplier. As a result, the accelerator with memory (14) generalizes the standard economic concepts of the accelerator and the multiplier [9].

### 3. Equation of Natural Growth with Memory Effect and its Solution

To take into account the fading memory effects in of natural growth model, we use the formula (14), which describes the relationship between the net investment and the marginal value of the output of order $\alpha > 0$ [6, 7, 8]. Substituting expressions (2) and (3) into equation (14), we obtain

$$(D_{0+}^\alpha Y)(t) - \frac{m \cdot (P-a)}{v} \cdot Y(t) = -\frac{m \cdot b}{v}. \tag{15}$$

Equation (15) is a fractional differential equation with fractional derivative of order $\alpha > 0$. Model of natural growth, which is based on equation (15), takes into account the effects of memory with power-law fading of order $\alpha \geq 0$. Equation (15) with $\alpha = 1$ gives equation (6), which describes a model of natural growth without memory effect.

We obtain the solution of equation (15) of the natural growth model that takes into account the effects of fading memory. It is known, [15, p. 323] that the fractional differential equation

$$(D_{0+}^\alpha Y)(t) - \lambda \cdot Y(t) = f(t), \tag{16}$$

where $f(\tau) \in C[0,t]$ is a real-valued function that is defined on the half-line t>0 with initial conditions $Y^{(k)}(0) = c_k$ (k=0,…,n-1), is solvable and has unique solution of the form

$$Y(t) = \int_0^t (t-\tau)^{\alpha-1} \cdot E_{\alpha,\alpha}[\lambda \cdot (t-\tau)^\alpha] \cdot f(\tau)d\tau +$$
$$\sum_{k=0}^{n-1} c_k \cdot t^k \cdot E_{\alpha,k+1}[\lambda \cdot t^\alpha], \tag{17}$$

where n-1<α≤n, $E_{\alpha,\beta}[z]$ is the two-parameter Mittag-Leffler function [15, p. 42], which is defined by the equation

$$E_{\alpha,\beta}(z) := \sum_{k=0}^{\infty} \frac{z^k}{\Gamma(\alpha k + \beta)}. \qquad (18)$$

The Mittag-Leffler function $E_{\alpha,\beta}(z)$ is a generalization of the exponential function $e^z$, since $E_{1,1}(z) = e^z$. It can be seen that equation (15) can be represented in the form (17) with

$$\lambda = \frac{m \cdot (P-a)}{v} \cdot, \quad f(t) = -\frac{m \cdot b}{v}. \qquad (19)$$

The solution of equation (15) has the form

$$Y(t) = \frac{-b}{(P-a)} \cdot \int_0^t (t-\tau)^{\alpha-1} \cdot E_{\alpha,\alpha}\left[\frac{m \cdot (P-a)}{v} \cdot (t-\tau)^\alpha\right] d\tau +$$
$$\sum_{k=0}^{n-1} Y^{(k)}(0) \cdot t^k \cdot E_{\alpha,k+1}\left[\frac{m \cdot (P-a)}{v} \cdot t^\alpha\right]. \qquad (20)$$

The calculation of the integral in equation (20) by using the change of variable ξ = t-τ, the definition (18) of the Mittag-Leffler function and term by term integration, gives solution (20) in the form

$$Y(t) = \frac{b}{(P-a)}\left(1 - E_{\alpha,1}\left[\frac{m \cdot (P-a)}{v} \cdot t^\alpha\right]\right) +$$
$$\sum_{k=0}^{n-1} Y^{(k)}(0) \cdot t^k \cdot E_{\alpha,k+1}\left[\frac{m \cdot (P-a)}{v} \cdot t^\alpha\right], \qquad (21)$$

where n-1<α≤n, and $Y^{(k)}(0)$ are the values of the derivatives of a function of the order of Y(t) at t = 0. Solution (21) describes the economic dynamics in the framework of natural growth model with power-law fading memory.

## 4. Solutions of Natural Growth Model with Memory

For 0<α≤1 (n=1) solution (21) has the form

$$Y(t) = \frac{b}{(P-a)}\left(1 - E_{\alpha,1}\left[\frac{m \cdot (P-a)}{v} \cdot t^\alpha\right]\right) + Y(0) \cdot E_{\alpha,1}\left[\frac{m \cdot (P-a)}{v} \cdot t^\alpha\right]. \qquad (22)$$

For α=1 expression (22) gives solution (6) since $E_{\alpha,1}[z] = e^z$, i.e. the solution of equation (21) with α = 1 is exactly the same as the standard solution of equations of the natural growth model without memory. Equation (22) can be rewritten as

$$Y(t) = \frac{b \cdot \Gamma(\alpha)}{(P-a)} + \left(Y(0) - \frac{b \cdot \Gamma(\alpha)}{(P-a)}\right) \cdot E_{\alpha,1}\left[\frac{m \cdot (P-a)}{v} \cdot t^\alpha\right]. \qquad (23)$$

If the condition $Y(0) - \frac{b}{(P-a)} > 0$ holds, then solution (23) describes the growth of output, and if $Y(0) - \frac{b}{(P-a)} < 0$, then (23) describes a downturn of output.

To illustrate the growth and downturn of output Y(t), which is described by equation (23) of the natural growth model with the fading memory at 0<α≤1, we give Figures 1-2.

The output Y(t), which is described by (23) of the natural growth model with the memory of fading order α = 0.9, is present on Figure 1 for Y(0)=12 and b=2, P-a=0.2, m=20, v=30. We can see that the memory effect with order α = 0.9 (α = 1 corresponds to processes without memory on graph 1 in Figure 1) slows the growth of output (graph 2 in Figure 1).

The output Y(t), which is described by (23) of the natural growth model with the memory of fading order α = 0.4, is present on Figure 2 for Y(0)=12 and b=2, P–a=0.2, m=20, v=30. We can see that the memory effect with order α = 0.5 leads to the strong slowing of output growth (graph 2 in Figure 2) in comparison with the model without memory effect (graph 1 in Figure 2).

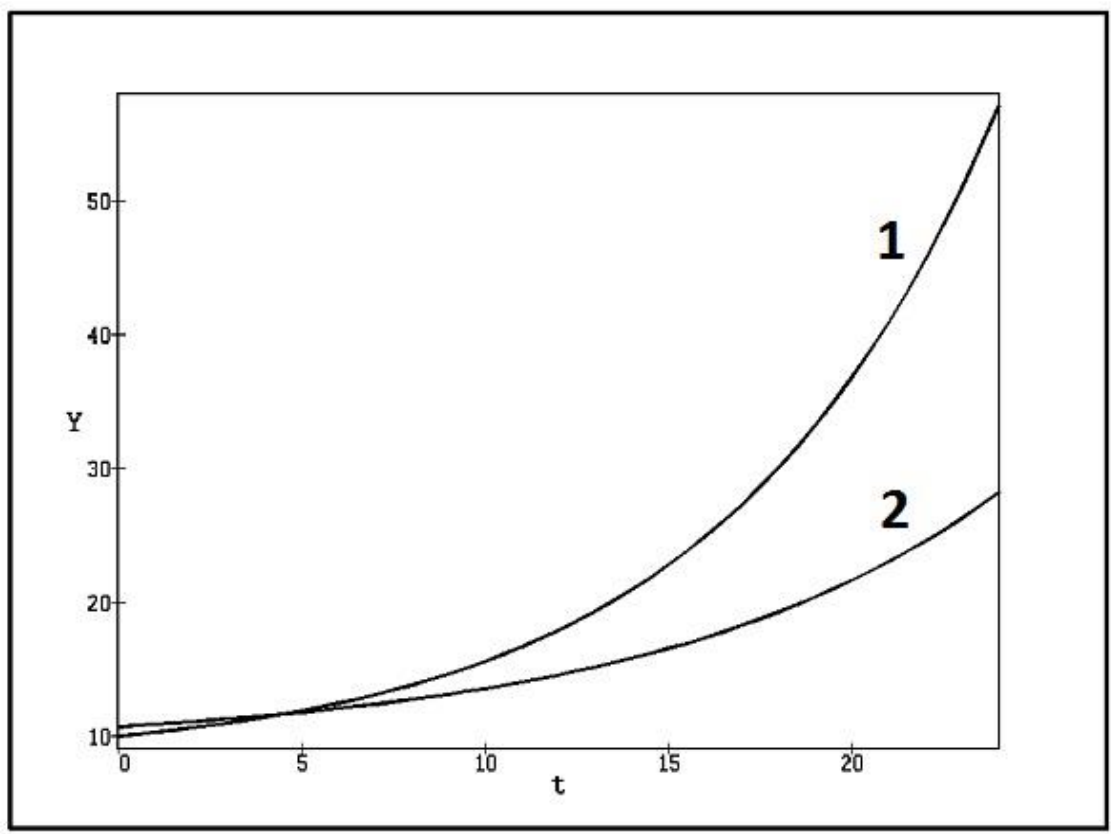

**Figure 1.** Memory effect leads to slower economic growth: Solution (6) of equation (4), which describe the natural growth model without memory effect (α=1), is present on graph 1. Solution (23) of equation (15), which describe natural growth model with memory of fading order α=0.9, is present on graph 2 for Y(0)=12 and b=2, P–a = 0.2, m=20, v=30, Y(0)=12.

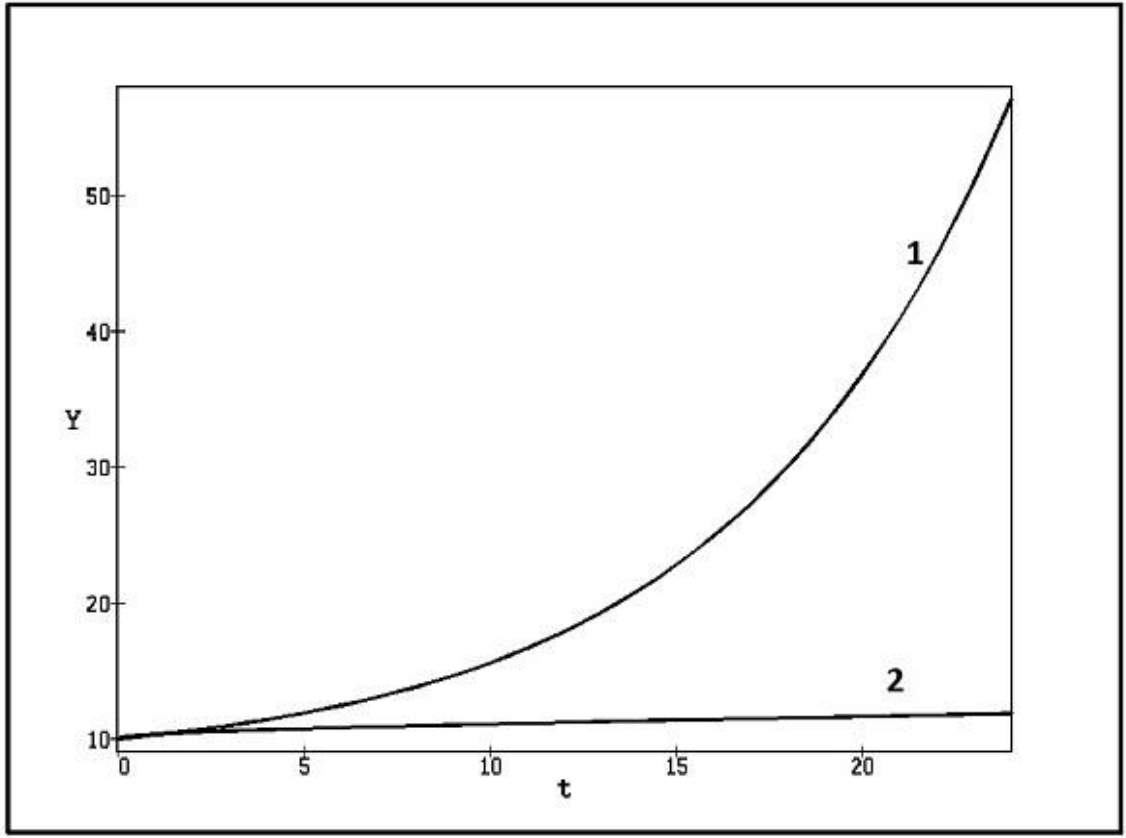

**Figure 2.** Memory effect leads to slower economic growth. The smaller value of parameter corresponds slower economic growth: Solution (6) of equation (4), which describe the natural growth model without memory effect (α=1), is present on graph 1. Solution (23) of equation (15), which describe natural growth model with memory of fading order α=0.4, is present on graph 2 for Y(0)=12 and b=2, P–a=0.2, m=20, v=30.

For $1 < \alpha \leq 2$ (n = 2), solution (21) has the form

$$Y(t) = \frac{b}{(P-a)}\left(1 - E_{\alpha,1}\left[\frac{m \cdot (P-a)}{v} \cdot t^\alpha\right]\right) + Y(0) \cdot E_{\alpha,1}\left[\frac{m \cdot (P-a)}{v} \cdot t^\alpha\right] +$$
$$Y^{(1)}(0) \cdot t \cdot E_{\alpha,2}\left[\frac{m \cdot (P-a)}{v} \cdot t^\alpha\right], \tag{24}$$

where $Y^{(1)}(0)$ is the value of the first order derivative of the function Y(t) at t = 0, i.e. the marginal value of output at the initial time.

To illustrate the growth and downturn of output Y(t), which is described by (24) of the natural growth model with the fading memory at $1<\alpha<2$, we give Figures 3-4. The output Y(t), which is described by (24) of the model with the memory of fading order α = 1.1, is present on Figure 3 for $Y^{(1)}(0) = 0.1$, Y(0)=12 and b=2, P–a=0.2, m=20, v=30. We can see that the memory effect with order α=1.1 leads to the acceleration of output growth (graph 2 in Figure 3) compared with the model without memory effect (graph 1 in Figure 3).

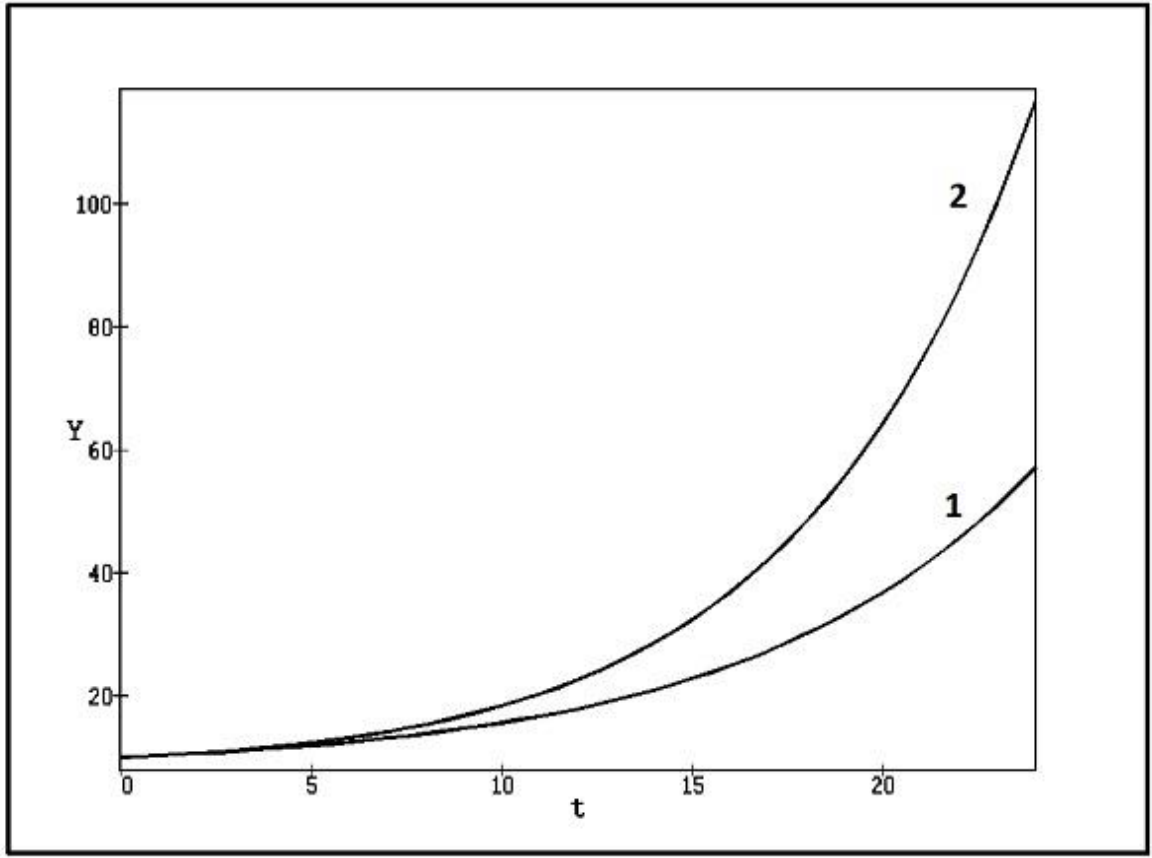

**Figure 3.** Memory effect leads to faster economic growth: Graph 1 presents solution (6) of equation (4) of the natural growth model without memory effect. Graph 2 presents solution (24) of equation (15) of the natural growth model with fading memory of order α = 1.1, where $Y^{(1)}(0) = 0.1$, Y(0)=12, and b=2, P–a=0.2, m=20, v=30.

The output Y(t), which is described by (24) of the natural growth model with the memory of fading order α = 1.1, is present on Figure 4 for $Y^{(1)}(0) = 10$, Y(0)=12, and b=4, P–a=0.2, m=20, v=35. We can see that the memory effect with order α = 1.1 can lead to the growth of output (graph 2 in Figure 4) instead of the downturn that is described by the model without memory effect (graph 1 in Figure 4).

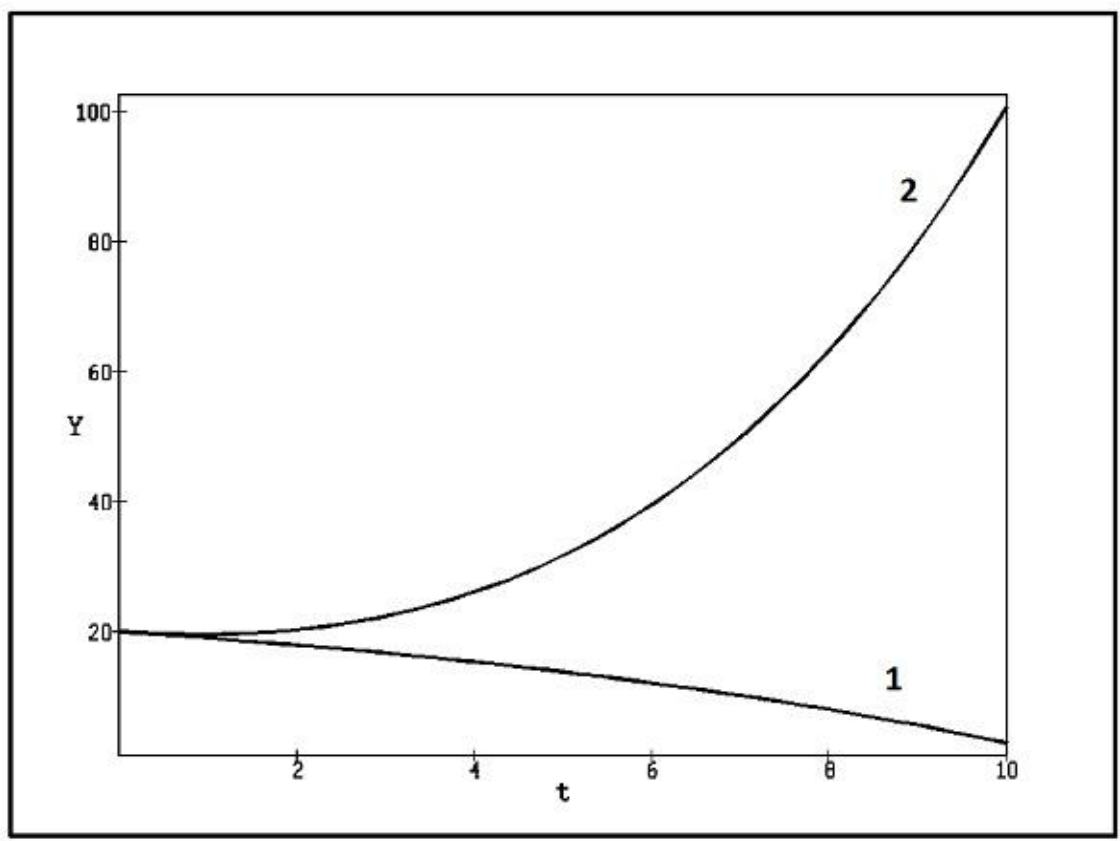

**Figure 4.** Memory effect leads to economic growth instead of downturn: Graph 1 presents solution (6) of equation (4) of the natural growth model without memory effect. Graph 2 presents solution (24) of equation (15) of the natural growth model with fading memory of order α = 1.1, where $Y^{(1)}(0) = 10$, Y(0)=12, and b=4, P–a=0.2, m=20, v=35.

Solutions (23) and (24) of the natural growth model with memory, which are present on Figures 1-4 for different values of memory fading, show that the value of the output strongly depends on the presence or absence of memory effect. The memory effect with fading order 0<α<1 can lead to the deceleration of output growth (graph 2 of Figures 1 and 2) compared with the output growth of model without memory (graph 1 of Figure 1 and Fig 2). The memory effect with fading order 1 <α <2 can lead to the output grows is accelerating (graph 2 of Figure 3) compared to the standard model without memory (graph 1 of Figure 3), or the output grows (graph 2 of Figure 4) instead of the output downturn of model without memory (graph 1 of Fig 4). As a result, we see the neglect of the memory effects in economic models may lead to qualitatively different results.

### 5. Conclusions

As a result, the memory effect with fading order 0<α<1 results in slower growth of output, compared with the standard model without memory. The memory effect with fading order 1<α<2 can give faster growth of output in comparison with the model without memory. Similar dependence of the economic dynamics on the memory effects arise in the of Harrod-Domar and Keynes models [18, 19, 20, 21].

The case of slower growth for $0 < \alpha < 1$, and accelerate growth for $1 < \alpha < 2$, is similar to the situation arising in physics in the description of anomalous diffusion [22], which is realized in the form of subdiffusion at $0 < \alpha < 1$ and superdiffusion at $1 < \alpha < 2$. The equations of anomalous diffusion have been used to describe financial processes [23, 24, 25, 26, 27, 28, 29].

In general, in economic models we should take into account the memory effects that are based on the fact that economic agents remember the story of changes of exogenous and endogenous variables that characterize the economic process. The proposed generalization of the simplest economic model (model of natural growth), which takes into account the effects of fading memory, has shown that the inclusion of memory effects can lead to significant changes in economic phenomena and processes.